\documentclass[a4paper,UKenglish,cleveref, autoref, thm-restate]{lipics-v2021}
\usepackage{packages}
\usepackage{macros}

\hideLIPIcs
\nolinenumbers

\usepackage{tikz}
\usetikzlibrary{arrows.meta} 
\usepackage{proofzilla}

\newemptyvertex{node}{}
\defedgetype{flow}{draw=blue}{}
\defedgetype{dirflow}{-Latex,draw=blue}{}
\defedgetype{flowtwo}{draw=red}{}
\defedgetype{dirflowtwo}{-Latex,draw=red}{}

\defedgetype{dirflowblue}{-Latex,draw=blue}{}
\defedgetype{dirflowred}{-Latex,draw=red}{}
\definecolor{black}{gray}{0}
\defedgetype{flowblack}{-Latex,draw=black}{} 

\title{Global Trace Condition for Injective Transition Graphs} 

\author{Stefano Berardi}
           {University of Turin, Turin, Italy}
           {stefano.berardi@unito.it}
           {}
           {}

\author{Ugo {de' Liguoro}}
           {University of Turin, Turin, Italy}
           {ugo.deliguoro@unito.it}
           {https://orcid.org/0000-0003-4609-2783}
           {}
           
\author{Daniel Osorio-Valencia}
           {University of Turin, Turin, Italy}
           {danielfabian.osoriovalencia@unito.it}
           {https://orcid.org/0009-0002-4820-1867}
           {}

\authorrunning{S. Berardi, U. de' Liguoro and D. Osorio-Valencia} 

\Copyright{S. Berardi, U. de' Liguoro and D. Osorio-Valencia} 



\category{} 

\relatedversion{} 

\begin{document}

\maketitle

\begin{abstract}
We formulate the {\em global trace condition}, GTC, for cyclic proofs as a particular case of the {\em size-change termination} condition or SCT. We express cyclic proofs through a variant of size-change graphs, called transition graphs, whose edges are labelled by (finite) bi-partite relations. 

We show that injective transition graphs suffice to encode cyclic pre-proofs for a variant of Brotherston and Simpson's sequent calculus with inductive definitions. 
The injectivity of transition graphs corresponds to the restriction of SCT to fan-in free graphs.
Then we prove that an essentially weaker condition implies SCT in this case, and hence GTC.
\end{abstract}

\section{Introduction}
\label{sec:Introduction}

Cyclic proof systems are the formalization of a particular way of reasoning with inductively defined predicates, where derivations using induction rules are replaced by infinite regular proof trees, whose infinite branches satisfy the ``infinite descent'' condition. This condition is inspired by Fermat’s method of infinite descent. If such a branch exists in a proof, then an infinite sequence of decreasing-size values of inductively defined formal parameters should exist, contradicting their well-founded definition. As a consequence, a cyclic proof encodes an underlying inductive argument that is sound.

This criterion,
 also called {\em Global Trace Condition} (GTC) in the literature, is the adaptation of Fermat's 
principle to cyclic proofs in sequent calculi, extended with Martin-Löf's  inductive definitions \cite{MartinLof}. It has been adopted by Brotherston and Simpson
\cite{Brotherston1, Brotherston2,BrotherstonSimpson}, where it implies the correctness of the 
conclusion of an infinite proof-tree \cite[Proposition 4.2.10]{Brotherston1}.
Counterintuitive as it may be, a good strategy 
to look for a finite proof is to look for a regular and \emph{infinite} proof satisfying GTC first.
Indeed, the use of GTC combines well both with proof search and with program synthesis
on realistic examples \cite{10.1007/978-3-642-35182-2_25,10.1145/3632888, 10.1007/978-3-031-90643-5_18,  10.1145/3453483.3454087,  10.1145/3018610.3018623, Stratulat_2021, Brotherston2020}. 

As observed in \cite{Brotherston1}, GTC is a termination argument. 
On the other hand, in \cite{Lee1} Lee, Jones, and Ben-Amram have
introduced the \emph{size-change termination} condition (SCT) 
to provide a sufficient condition for the totality of partial recursive maps on natural numbers,
initiating a thread of research on program termination; see \cite{Lee2,Ben-AmramLee2007,Lee3} and references there.
A major result in \cite{Lee1} is that deciding that SCT holds for a set of graphs, called size-change graphs, is a PSPACE-complete problem.

In her recent study \cite{Mirai}, Ikebuchi has shown a deep relationship among GTC and SCT, by proving that a set of size-change graphs can be associated with each cyclic pre-proof so that it satisfies GTC if and only if the graphs associated with it satisfy SCT.

In this paper, we go further in investigating the relation between cyclic proofs and SCT. 
In \cite{Lee2,Ben-AmramLee2007,Lee3}, Lee and Ben-Amram have explored the complexity of the problem of deciding the SCT condition by
considering some particular cases of size-change graphs. In the case of fan-in free graphs, namely when the in-degree of all nodes is at most one, algorithms solving SCT problem have an exponential lower bound given by the size of ranking functions (roughly related to the size of the graph), in place of the triple exponentials needed in the general case. By considering a sequent calculus in the original sense of Gentzen \cite{Gentzen1969}, where sequents are pairs of (finite) lists instead of multisets or sets of formulas as in Brotherston and Simpson systems, we show that the notion of trace in a pre-proof can be identified with a thread in a set of fan-in free size-change graphs, while preserving the equivalence among GTC and SCT established by Ikebuchi.

Technically, this is obtained by using a variant of the notion of size-change graphs, which we call {\em transition graphs}. These are directed and cyclic graphs, rather than multigraphs, where edges are labeled by certain relations, modeling the \quotationMarks{stationary} and 
 \quotationMarks{progressing} parameters in a proof. We adopt transition graphs instead of (sets of) size-change graphs because they are more natural when modeling the GTC of cyclic proofs, where paths represent the branches in the proof.

In our formalism, the fan-in free condition amounts to restricting the relations to the injective case, where two distinct parameters cannot be related to the same third one. The existence of a path in a transition graph determines a relation which is the composition of those labeling its edges, which allows us to define a new condition called here the {\em Progress Domain Condition} 
(PDC) solely in terms of the domain of the relations labeling cycles in the transitive closure of a transition graph. 

We show that PDC is an essentially weaker condition than SCT, and therefore than GTC, but it is equivalent to them in the case of transition graphs with only injective relations associated with their edges, namely in the case of fan-in free size-change graphs, and therefore it can be used to discriminate valid cyclic pre-proofs in our proof system. Also, we think that deciding PDC should not be harder than deciding SCT in the case of fan-in free size-change graphs, and we conjecture it might be a faster exponential 
than the measure based on the size of ranking functions in \cite{Lee3}.

\medskip
The paper is organized as follows: in Section \ref{sec:cyclic} we introduce our Gentzen-style version of the systems $\CLKIDomega$ and $\CLJIDomega$; in Section
\ref{sec:size-change} we define the transition graphs and SCT, which are paired with cyclic pre-proofs in Section \ref{sec:transition-graph-proofs}. The main results of the paper are in Section \ref{sec:progressDomainCond}, where the PDC condition is studied in the case of injective transition graphs. Then in Section \ref{sec:Related} we
discuss the related works and conclude.

\section{Cyclic Proofs and the Global Trace Condition}
\label{sec:cyclic}

In this section, we define the GTC, for the cyclic proof systems $\CLKIDomega$ and $\CLJIDomega$. The difference with the previous formulation is that we consider an injective trace relation, as we will see.

$\CLKIDomega$ and $\CLJIDomega$ are both systems for first-order theories, hence we consider a signature $\Sigma$, which contains symbols for constants, functions, and predicates of individuals. Terms, formulas, and $\Sigma$-structures are defined as usual. We consider sequents as originally defined by Gentzen using finite lists, instead of the formulations in \cite{Brotherston2, BrotherstonSimpson} using sets or multisets. 

In the case of $\CLKIDomega$, sequents have the form $\Gamma \vdash \Delta$, while in $\CLJIDomega$, they are defined as $\Gamma \vdash \phi$, where $\Gamma$ and $\Delta$ are finite lists of formulas, and $\phi$ is either a single formula or the empty list. 
The comma between lists of formulas (e.g. $\Gamma, \Delta$) denotes list concatenation. For any formula $F$ in a list $\Gamma$, we define the \emph{occurrence} of $F$ in $\Gamma$ as the formula $F$ paired with the position $i$ where it appears in $\Gamma$, written as $(F,i)$. Predicate symbols are divided between \emph{inductive predicates}, as in Martin-Löf's \emph{productions} \cite{MartinLof}, and \emph{ordinary predicates}. To formally define inductive predicates, we introduce the notion of \emph{inductive definition set}.

\begin{definition}[Inductive definition set]
    \label{defIndSet}
    An inductive definition set $\Phi$ is a finite set of \emph{productions} $r$ of the form:

    \begin{center}
        \centering
        \begin{prooftree}
            \hypo{Q_1\mathbf{u_1} \ldots Q_h\mathbf{u_h}}
            \hypo{P_1\mathbf{t_1} \ldots P_m\mathbf{t_m}}
            \infer2[r]{P\mathbf{t}}
        \end{prooftree}
    \end{center}

\noindent
where $Q_1, \ldots, Q_h$ are ordinary predicate symbols, $P_1, \ldots, P_m$ are inductive predicate symbols, and $\mathbf{u_1}, \ldots, \mathbf{u_h}$, $\mathbf{t_1}, \ldots, \mathbf{t_m}$, $\mathbf{t}$ are tuples of terms.

A production of conclusion $P \mathbf{t}$ for some $\mathbf{t}$ is said to be a production for $P$. A formula is said \emph{inductive} when it is an atomic formula $P\mathbf{t}$ for some inductive predicate $P$ and some $\mathbf{t}$.
\end{definition} 

\begin{example}
\label{ex:NEO}
We define the inductive definition set $\Phi_0 = \{r_1, r_2,r_3,r_4,r_5\}$, 
using the following production rules:

\begin{center}
    \centering
    \begin{minipage}{0.10\textwidth}
        \centering
        \begin{prooftree}
            \hypo{}
            \infer1[$r_1$]{N0}
        \end{prooftree}
    \end{minipage}
    \begin{minipage}{0.10\textwidth}
        \centering
        \begin{prooftree}
            \hypo{Nx}
            \infer1[$r_2$]{Nsx}
        \end{prooftree}
    \end{minipage}
    \begin{minipage}{0.10\textwidth}
        \centering
        \begin{prooftree}
            \hypo{}
            \infer1[$r_3$]{E0}
        \end{prooftree}
    \end{minipage}
    \begin{minipage}{0.15\textwidth}
        \centering
        \begin{prooftree}
            \hypo{Ox}
            \infer1[$r_4$]{Esx}
        \end{prooftree}
    \end{minipage}
    \begin{minipage}{0.12\textwidth}
        \centering
        \begin{prooftree}
            \hypo{Ex}
            \infer1[$r_5$]{Osx}
        \end{prooftree}
    \end{minipage}
\end{center}

Here $0$ is a constant representing the number zero and $s$ is a unary function representing the successor function. The predicate $N$ is used to inductively represent the natural numbers, 
$E$ and $O$ are \emph{mutually defined} inductive predicates defining even and odd numbers respectively.
\end{example}

\begin{figure}


\begin{mdframed}
\adjustbox{scale=0.9}
{
\begin{tabular}{ccc}


  \begin{minipage}{0.3\textwidth}
    \centering
    \begin{prooftree}
      \hypo{}
      \infer1[(Axiom)]{\Gamma, F \vdash \Delta, F  } 
    \end{prooftree}
  \end{minipage} &
  \begin{minipage}{0.3\textwidth}
    \centering
    \begin{prooftree}
        \hypo{\vnode1{\Gamma} \vdash \vnode2{\Delta}}
        \infer1[(WeakL)]{\vnode3{\Gamma}, F \vdash \vnode4{\Delta}} 
    \end{prooftree}

    $
    \dirflowblueedges{node3/node1} 
    $
  \end{minipage} &
  \begin{minipage}{0.3\textwidth}
    \centering
    \begin{prooftree}
        \hypo{\vnode1{\Gamma} \vdash \vnode2{\Delta}}
        \infer1[(WeakR)]{\vnode3{\Gamma} \vdash \vnode4{\Delta}, F} 
    \end{prooftree}

    $
    \dirflowblueedges{node3/node1} 
    $
  \end{minipage}
  
  \\[1em]

  \begin{minipage}{0.3\textwidth}
    \centering
    \begin{prooftree}
        \hypo{\vnode1{\Gamma}, \vnode2{F}, \vnode3{F} \vdash \vnode4{\Delta}}
        \infer1[(ContrL)]{\vnode5{\Gamma}, \vnode6{F} \vdash \vnode7{\Delta}} 
    \end{prooftree}

    $
    \dirflowblueedges{node5/node1, node6/node2, node6/node3}
    $
  \end{minipage} &
  \begin{minipage}{0.3\textwidth}
    \centering
    \begin{prooftree}
        \hypo{\vnode1{\Gamma} \vdash \vnode2{F}, \vnode3{F}, \vnode4{\Delta}}
        \infer1[(ContrR)]{\vnode5{\Gamma} \vdash \vnode6{F}, \vnode7{\Delta}} 
    \end{prooftree}

    $
    \dirflowblueedges{node5/node1}
    $
  \end{minipage} &
  \begin{minipage}{0.3\textwidth}
    \centering
    \begin{prooftree}
        \hypo{\vnode1{\Gamma} \vdash F, \vnode2{\Delta}}
        
        \hypo{F, \vnode3{\Gamma'} \vdash \vnode4{\Delta'}}

        \infer2[(Cut)]{\vnode5{\Gamma}, \vnode6{\Gamma'} \vdash \vnode7{\Delta}, \vnode8{\Delta'}} 
    \end{prooftree}

    $
    \dirflowblueedges{node5/node1, node6/node3}
    $
  \end{minipage}
  \\[1em]

  \begin{minipage}{0.3\textwidth}
    \centering
    \begin{prooftree}
        \hypo{\vnode1{\Gamma} \vdash \vnode2{\Delta}}
        \infer1[(Subst)]{\vnode3{\Gamma[\theta]} \vdash \vnode4{\Delta[\theta]}} 
    \end{prooftree}

    $
    \dirflowblueedges{node3/node1} 
    $
  \end{minipage} &
  \begin{minipage}{0.3\textwidth}
    \centering
    \begin{prooftree}
        \hypo{\vnode1{\Gamma}, \vnode2{F}, \vnode3{G}, \vnode4{\Gamma'} \vdash \vnode5{\Delta}}
        \infer1[(ExL)]{\vnode6{\Gamma}, \vnode7{G}, \vnode8{F}, \vnode9{\Gamma'} \vdash \vnode0{\Delta}} 
    \end{prooftree}

    $
    \dirflowblueedges{node6/node1, node9/node4}
    \bentdirflowblueedges{node7/node3/-15, node8/node2/15} 
    $
  \end{minipage} &
  \begin{minipage}{0.3\textwidth}
    \centering
    \begin{prooftree}
        \hypo{\vnode1{\Gamma} \vdash \vnode2{\Delta}, \vnode3{F}, \vnode4{G}, \vnode5{\Delta'}}
        \infer1[(ExR)]{\vnode6{\Gamma}\vdash \vnode7{\Delta}, \vnode8{G}, \vnode9{F}, \vnode0{\Delta'}} 
    \end{prooftree}

    $
    \dirflowblueedges{node6/node1}
    $
  \end{minipage}
  \\[1em]

  \begin{minipage}{0.3\textwidth}
    \centering
    \begin{prooftree}
          \hypo{\vnode1{\Gamma} \vdash \vnode2{F}, \vnode3{\Delta}}
          \infer1[($\neg$L)]{\vnode4{\Gamma}, \vnode5{\neg F} \vdash \vnode6{\Delta}} 
      \end{prooftree} 

      $
      \dirflowblueedges{node4/node1}
      $
  \end{minipage} &
  \begin{minipage}{0.3\textwidth}
    \centering
    \begin{prooftree}
          \hypo{\vnode1{\Gamma}, \vnode2{F} \vdash \vnode3{\Delta}}
          \infer1[($\neg$R)]{\vnode4{\Gamma} \vdash \vnode5{\neg F} \vnode6{\Delta}} 
      \end{prooftree} 

      $
      \dirflowblueedges{node4/node1}
      $
  \end{minipage} &
  \begin{minipage}{0.3\textwidth}
    \centering
    \begin{prooftree}
          \hypo{\vnode1{\Gamma}, \vnode2{F}, \vnode3{G} \vdash \vnode4{\Delta}}
          \infer1[($\land$L)]{\vnode5{\Gamma}, \vnode6{F \land G} \vdash \vnode7{\Delta}} 
      \end{prooftree} 

      $
      \dirflowblueedges{node5/node1}
      $
    
  \end{minipage}
  \\[1em]

  \begin{minipage}{0.3\textwidth}
    \centering
      \begin{prooftree}
          \hypo{\vnode1{\Gamma} \vdash \vnode2{F}, \vnode3{\Delta}}

          \hypo{\vnode4{\Gamma} \vdash \vnode5{G}, \vnode6{\Delta}}

          \infer2[($\land$R)]{\vnode7{\Gamma} \vdash \vnode8{F \land G}, \vnode9{\Delta}} 
      \end{prooftree} 

      $
      \dirflowblueedges{node7/node1, node7/node4}
      $
  \end{minipage} &
  \begin{minipage}{0.3\textwidth}
    \centering
    
      \begin{prooftree}
          \hypo{\vnode1{\Gamma}, \vnode2{F} \vdash \vnode3{\Delta}}

          \hypo{\vnode4{\Gamma}, \vnode5{G} \vdash \vnode6{\Delta}}

          \infer2[($\lor$L)]{\vnode7{\Gamma}, \vnode8{F \lor G} \vdash \vnode9{\Delta}} 
      \end{prooftree} 

      $
      \dirflowblueedges{node7/node1, node7/node4}
      $
    
  \end{minipage} &
  \begin{minipage}{0.3\textwidth}
    \centering
    
    \begin{prooftree}
          \hypo{\vnode1{\Gamma} \vdash \vnode2{F}, \vnode3{G}, \vnode4{\Delta}}
          \infer1[($\lor$R)]{\vnode5{\Gamma} \vdash \vnode6{F \lor G}, \vnode7{\Delta}} 
      \end{prooftree} 

      $
      \dirflowblueedges{node5/node1}
      $
    
  \end{minipage}
  \\[1em]

  \begin{minipage}{0.3\textwidth}
    \centering
    
      \begin{prooftree}
          \hypo{\vnode1{\Gamma} \vdash \vnode2{F}, \vnode3{\Delta}}

          \hypo{\vnode4{\Gamma}, \vnode5{G} \vdash \vnode6{\Delta}}

          \infer2[($\rightarrow$L)]{\vnode7{\Gamma}, \vnode8{F \rightarrow G} \vdash \vnode9{\Delta}} 
      \end{prooftree} 

      $
      \dirflowblueedges{node7/node1, node7/node4}
      $
    
  \end{minipage} &
  \begin{minipage}{0.3\textwidth}
    \centering
    
    \begin{prooftree}
          \hypo{\vnode1{\Gamma}, \vnode2{F} \vdash \vnode3{G}, \vnode4{\Delta}}
          \infer1[($\rightarrow$R)]{\vnode5{\Gamma} \vdash \vnode6{F \rightarrow G}, \vnode7{\Delta}} 
      \end{prooftree} 

      $
      \dirflowblueedges{node5/node1}
      $
    
  \end{minipage} &
  \begin{minipage}{0.3\textwidth}
    \centering
    
    \begin{prooftree}
        \hypo{\vnode1{\Gamma[u/x,t/y]} \vdash \vnode2{\Delta[u/x,t/y]}}
        \infer1[(=L)]{\vnode3{\Gamma[t/x,u/y]}, t = u \vdash \vnode4{\Delta[t/x,u/y]}} 
    \end{prooftree}

    $
    \dirflowblueedges{node3/node1} 
    $
    
  \end{minipage}
  \\[1em]

  \begin{minipage}{0.3\textwidth}
    \centering
    
    \begin{prooftree}
      \hypo{}
      \infer1[(=R)]{\Gamma \vdash t = t,\Delta} 
    \end{prooftree}
    
  \end{minipage} &
  \begin{minipage}{0.3\textwidth}
    \centering
    
    \begin{prooftree}
        \hypo{\vnode1{\Gamma}, \vnode2{F[t/x]} \vdash \vnode3{\Delta}}
        \infer1[($\forall$L)]{\vnode4{\Gamma}, \vnode5{\forall x F} \vdash \vnode6{\Delta}} 
    \end{prooftree}

    $
    \dirflowblueedges{node4/node1}
    $
    
  \end{minipage} &
  \begin{minipage}{0.3\textwidth}
    \centering
    
    \begin{prooftree}
        \hypo{\vnode1{\Gamma} \vdash \vnode2{F[t/x]}, \vnode3{\Delta}}
        \infer1[($\exists$R)]{\vnode4{\Gamma} \vdash \vnode5{\exists x F}, \vnode6{\Delta}} 
    \end{prooftree}

    $
    \dirflowblueedges{node4/node1}
    $
    
  \end{minipage}
  \\[1em]

  \multicolumn{3}{c}{
  \begin{tabular}{cc}
    \begin{minipage}{0.4\textwidth}
    \centering
    
    \begin{prooftree}
        \hypo{\vnode1{\Gamma} \vdash \vnode2{F}, \vnode3{\Delta}}
        \infer1[$x \notin FV(\Gamma,\Delta)$ ($\forall$R)]{\vnode4{\Gamma} \vdash \vnode5{\forall x F}, \vnode6{\Delta}} 
    \end{prooftree}

    $
    \dirflowblueedges{node4/node1}
    $
    
    \end{minipage}

    &

    \begin{minipage}{0.4\textwidth}
    \centering
    
    \begin{prooftree}
        \hypo{\vnode1{\Gamma}, \vnode2{F} \vdash \vnode3{\Delta}}
        \infer1[$x \notin FV(\Gamma,\Delta)$ ($\exists$L)]{\vnode4{\Gamma}, \vnode5{\exists x F} \vdash \vnode6{\Delta}} 
    \end{prooftree}

    $
    \dirflowblueedges{node4/node1}
    $
    
    \end{minipage}
  \end{tabular}
}
\end{tabular}
}
\end{mdframed}
 
  \caption{Rules of LK. Blue arrows define the trace relation between 
  atomic inductive formulas (Def. \ref{def:subOccur}).}
   \label{LKrules}
\end{figure}

We define the proof system $\CLKIDomega$ for inductive predicates as an extension of the classical logic LK, whose rules are reported in Figure \ref{LKrules}, with additional rules used to handle inductive predicates on the right and left part of the sequent. These rules for inductive predicates are defined as follows: let $\Phi = \{r_1, \ldots, r_n\}$ be the inductive definition set, such that all $r_i$ are productions as in Definition \ref{defIndSet}.

\begin{itemize}
    \item For each production $r_i$ for an inductive predicate $P$, the system $\CLKIDomega$ includes a corresponding right introduction rule for $P$, denoted ($P\text{R}_i$):
    
    \begin{equation}
    \label{def:caseRuleRight}
    \begin{prooftree}
        \hypo{\vnode1{\Gamma} \vdash \vnode2{Q_1\mathbf{u_1}}, \vnode3{\Delta} \ldots \vnode4{\Gamma} \vdash \vnode5{Q_h\mathbf{u_h}}, \vnode6{\Delta}}
        \hypo{\vnode7{\Gamma} \vdash \vnode8{P_1\mathbf{t_1}}, \vnode9{\Delta} \ldots \vnode{10}{\Gamma} \vdash \vnode{11}{P_m\mathbf{t_m}}, \vnode{12}{\Delta}}
        \infer2[($P\text{R}_i$)]{\vnode{13}{\Gamma} \vdash \vnode{14}{P\mathbf{t}}, \vnode{15}{\Delta}}
    \end{prooftree}
    $
    \bentdirflowblueedges{node13/node1/20, node13/node4/20, node13/node7/0, node13/node10/-30}
    $
    \end{equation}

    \item For all inductive predicates $P$ there is a \emph{single} left introduction rule associated to $P$ and to all productions $r_i  \in \Phi$ for $P$. It is denoted \emph{case split rule for $P$} or (Case $P$):
    
    \begin{equation}
    \label{def:caseRuleLeft}
    \begin{prooftree}
        \hypo{ \{ \vnode1{\Gamma}, \vnode{11}{\mathbf{u} = \mathbf{t^i}}, \vnode3{Q^i_1\mathbf{u^i_1}}, \ldots , \vnode4{Q^i_h\mathbf{u^i_h}}, \vnode5{P^i_1\mathbf{t^i_1}}, \ldots, \vnode6{P^i_m\mathbf{t^i_m}} \vdash \vnode7{\Delta} \}_{1 \leq i \leq n} }
        \infer1[$(Case P)$]{\vnode8{\Gamma}, \vnode9{P\mathbf{u}} \vdash \vnode0{\Delta}}
    \end{prooftree}
    $
    \bentdirflowblueedges{node8/node1/20}
    \bentdirflowrededges{node9/node5/0, node9/node6/0} 
    $
    \end{equation}

    In the rule (Case $P$), for each production $r_i$, there is a corresponding sequent in the premises of the form $\Gamma, \mathbf{u} = \mathbf{t^i}, Q^i_1\mathbf{u^i_1}, \ldots , Q^i_h\mathbf{u^i_h}, P^i_1\mathbf{t^i_1}, \ldots, P^i_m\mathbf{t^i_m} \vdash \Delta$ subject to the restriction that any free variable in $\mathbf{u^i_1},...,\mathbf{u^i_h},\mathbf{t^i_1},...,\mathbf{t^i_m}$, or $\mathbf{t^i}$, does not occur free in $\Gamma, \Delta$ and $\mathbf{u}$.

\end{itemize}

To the usual presentation of these rules, we add a \emph{trace relation} among occurrences of inductive formulas, which is represented by {\color{blue} \bf blue} and {\color{red} \bf red} arrows from the conclusion to the premises, according to the clauses below and Definition \ref{def:subOccur}.

\begin{enumerate}[i)]
    \item Arrows of either color connect occurrences $(P\mathbf{u}, i)$ on the left side of the conclusion's sequent, to occurrences $(Q\mathbf{t}, j)$ on the left side of the premises' sequents of each rule, where $P$ and $Q$ are inductive predicates.
    \item The {\color{blue} \bf blue} arrow from a list $\Gamma$ in the conclusion to the same list $\Gamma$ in a premise \emph{defines a bijection} between occurrences $(P\mathbf{u},i)$ in the list $\Gamma$ in the conclusion, and $(P\mathbf{u},i)$ in the \emph{same position} of the list $\Gamma$ in the premise.
    \item The {\color{blue} \bf blue} arrow from a list $\Gamma[\theta]$ in the conclusion to $\Gamma$ in a premise, in the rule $(Subst)$, defines a bijection between occurrences $(P\mathbf{u}[\theta],i)$ in the list $\Gamma[\theta]$ in the conclusion, and $(P\mathbf{u},i)$ in the list $\Gamma$ in the premise.
    \item The {\color{blue} \bf blue} arrow from a list $\Gamma[t/x,u/y]$ in the conclusion to $\Gamma[u/x,t/y]$ in a premise, in the rule (=L), defines a bijection between occurrences $(P\mathbf{u}[t/x,u/y],i)$ in the list $\Gamma[t/x,u/y]$ in the conclusion, and $(P\mathbf{u}[u/x,t/y],i)$ in the list $\Gamma[u/x,t/y]$ in the premise.   
\end{enumerate}

\begin{definition}[Trace Relation]
    \label{def:subOccur}
    Let $r$ be a rule in $\CLKIDomega$, with conclusion $\Gamma \vdash \Delta$ and premises $\Gamma_1 \vdash \Delta_1, \ldots, \Gamma_n \vdash \Delta_n$. Moreover, consider the occurrences $(P\mathbf{u},i)$ in $\Gamma$ and $(Q\mathbf{t},j)$ in $\Gamma_k$ for some $ 1 \leq k \leq n$, where $P$ and $Q$ are inductive predicates. We say that $(P\mathbf{u},i)$ is connected to $(Q\mathbf{t},j)$ in $r$, noted $(P\mathbf{u},i) \rightarrow_{r} (Q\mathbf{t},j)$ if and only if there is an arrow drawn from $(P\mathbf{u},i)$ to $(Q\mathbf{t},j)$ in the definition of rule $r$ in the Figure \ref{LKrules} or the Formulas (\ref{def:caseRuleRight}) and (\ref{def:caseRuleLeft}). $(P\mathbf{u},i)$ is called a \emph{progressing} occurrence, if only {\color{red} \bf red} arrows originate from it, and a \emph{stationary} occurrence if all arrows out of it are {\color{blue} \bf blue}.
\end{definition}

The intended meaning of {\color{red} \bf red} progressing arrows and of {\color{blue} \bf blue} stationary arrows is: when $P\mathbf{t}$ progresses to $Q\mathbf{u}$ then the vector denoted by $\mathbf{t}$ has larger size than the vector denoted by $\mathbf{u}$, when there is a stationary connection then $\mathbf{t}$ and $\mathbf{u}$ denote the same vector.

\begin{figure}
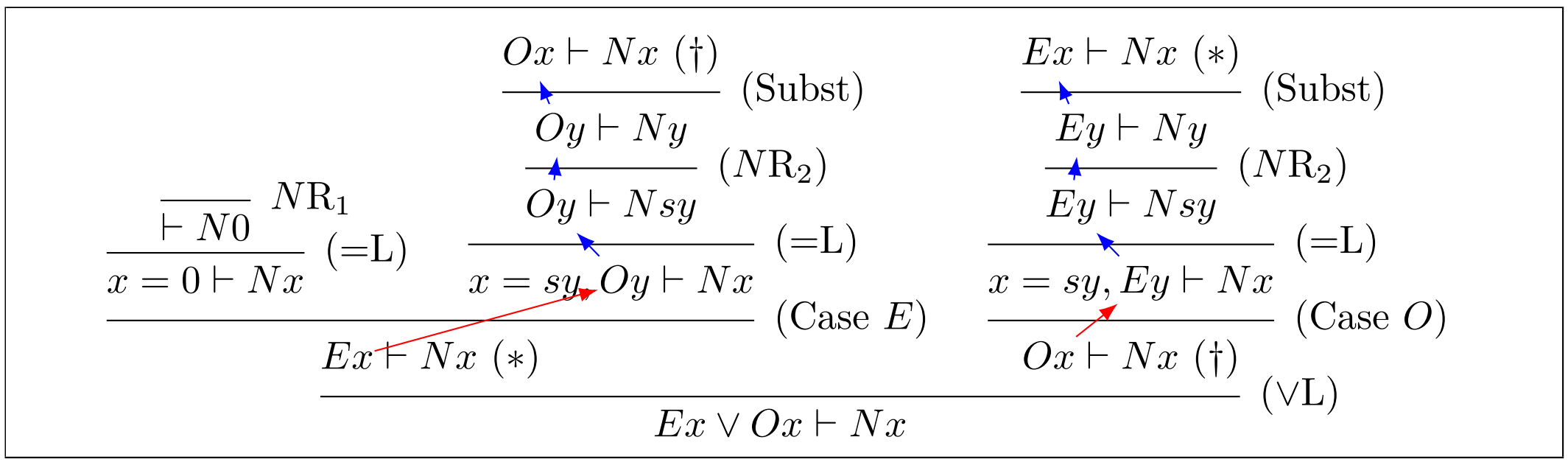


\begin{mdframed}
\centering
\adjustbox{scale=1}{
\begin{prooftree}

    \hypo{}
    \infer1[$N\text{R}_1$]{\vdash N0}
    \infer1[($=$L)]{x = 0 \vdash Nx}

    \hypo{\vnode0{Ox} \vdash Nx \ (\vnode{12}{\dag})}
    \infer1[(Subst)]{\vnode9{Oy} \vdash Ny}
    \infer1[($N\text{R}_2$)]{\vnode8{Oy} \vdash Nsy}
    \infer1[($=$L)]{x = sy, \vnode7{Oy} \vdash Nx}

    \infer2[(Case $E$)]{\vnode6{Ex} \vdash Nx \ (\vnode{11}{*})}

    \hypo{\vnode5{Ex} \vdash Nx \ (\vnode{10}{*})}
    \infer1[(Subst)]{\vnode4{Ey} \vdash Ny}
    \infer1[($N\text{R}_2$)]{\vnode3{Ey} \vdash Nsy}
    \infer1[($=$L)]{x = sy, \vnode2{Ey} \vdash Nx}
    \infer1[(Case $O$)]{\vnode1{Ox} \vdash Nx \ (\vnode{13}{\dag})}    

    \infer2[($\lor$L)]{Ex \lor Ox \vdash Nx}
\end{prooftree}

    $
    \dirflowblueedges{node2/node3, node3/node4, node4/node5, node7/node8, node8/node9, node9/node0}
    \dirflowrededges{node1/node2, node6/node7}
    $

}
\end{mdframed}

\caption{$\CLKIDomega$ cyclic pre-proof of the sequent $Ex \lor Ox \vdash Nx$. The function $\mathcal{F}$ is represented by labeling each bud and its corresponding companion node with the $(\dag)$ and $(*)$ symbols. 
The arrows from the trace relation are used to show the validating trace 
(see Def. \ref{def:traceCyclic}).}
\label{exampleCircProof}
\end{figure}

\begin{example}
\label{example-NEO-sequent}
For the inductive predicates defined in Example \ref{ex:NEO}, we add the following rules (for simplicity, we omit the rules that are not used in the proof of Figure \ref{exampleCircProof}):    

\begin{center}
    \centering
    \begin{minipage}{0.3\textwidth}
        \centering
        \adjustbox{scale=1}{
        \begin{prooftree}
            \hypo{}
            \infer1[($N\text{R}_1$)]{\Gamma \vdash N0, \Delta}
        \end{prooftree}
        }
    \end{minipage}
    \hfill
    \begin{minipage}{0.3\textwidth}
        \centering
        \adjustbox{scale=1}{
        \begin{prooftree}
            \hypo{\vnode2{\Gamma} \vdash Nx, \Delta}
            \infer1[($N\text{R}_2$)]{\vnode1{\Gamma} \vdash Nsx, \Delta}
        \end{prooftree}
        $
        \bentdirflowblueedges{node1/node2/15}
        $
        }
    \end{minipage}
    \hfill
    \begin{minipage}{0.3\textwidth}
        \centering
        \adjustbox{scale=1}{
        \begin{prooftree}
            \hypo{\vnode2{\Gamma}, x = sz, \vnode4{Ez} \vdash \Delta}
            \infer1[(Case $O$)]{\vnode1{\Gamma}, \vnode3{Ox} \vdash \Delta}
        \end{prooftree}
         $
        \bentdirflowblueedges{node1/node2/30}
        \dirflowrededges{node3/node4}
        $
        }
    \end{minipage}
    \hfill
    \\[1em]

    \begin{centering}
        \adjustbox{scale=1}{
        \begin{prooftree}
            \hypo{\vnode0{\Gamma}, x = 0 \vdash \Delta}
            \hypo{\vnode2{\Gamma}, x = sz,\vnode4{Oz} \vdash \Delta}
            \infer2[(Case $E$)]{\vnode1{\Gamma}, \vnode3{Ex} \vdash \Delta}
        \end{prooftree}
         $
         \bentdirflowblueedges{node1/node0/30}
        \bentdirflowblueedges{node1/node2/30}
        \dirflowrededges{node3/node4}
        $
        }
    \end{centering}

\end{center}

\end{example}

The proof system for $\CLJIDomega$ and its trace relation is naturally obtained by restricting the rules for inductive predicates over intuitionistic sequents, preserving the exchange rule, and replacing LK with its intuitionistic version, named LJ (see, for instance \cite{Berardi}). 

Our presentation differs from the previous one in that we consider an \emph{injective} trace
relation, as the next Proposition says.

\newcommand{\codom} {{\tt cod}}

\begin{proposition}
\label{prop:injective-trace}
\begin{enumerate}
\item
\label{prop:injective-trace-01}
Assume ${\mathcal R} = \{R_i | i \in I\}$ is a set of injective relations on $X \times Y$
with \emph{disjoint codomains}, namely: $\forall i,j\in I. (\exists x \in \codom(R_i) \cap \codom(R_j)) \implies i=j$.
Then $\cup {\mathcal R}$ is an injective relation on $X \times Y$. 
\item
\label{prop:injective-trace-02}
The trace relation in Def. \ref{def:subOccur} is injective.
\end{enumerate}
\end{proposition}

\begin{proof}
\begin{enumerate}
\item
Let $R = \cup {\mathcal R}$. Assume $x R y$ and $x' R y$ in order to prove that
$x = x'$. From $x R y$ and $x' R y$ we deduce
$x R_i y$ and $x' R_j y$ for some $i,j \in I$. From $y \in \codom(R_i) \cap \codom(R_j)$
we deduce that $i = j$, therefore $x R_i y$ and $x' R_i y$. 
From $R_i$ injective we conclude that $x = x'$.
\item
For all rules we considered a trace relation union of injective relations, 
on sequents and on single formulas, which are injective and have \emph{disjoint codomain}. 
By point \ref{prop:injective-trace-01} above, these unions define an injective trace relation.
\end{enumerate}
\end{proof}

\emph{Cyclic pre-proofs}, defined below, are used to formally represent a 
\emph{possibly infinite regular derivation} in $\CLKIDomega$ and $\CLJIDomega$
 through a \emph{finite graph} with a bud/companion relationship used to represent 
 the infinite tree.

\begin{definition}[Cyclic pre-proof] A $\CLKIDomega$ (resp. $\CLJIDomega$) pre-proof of sequent $\Gamma \vdash \Delta$ (resp. $\Gamma \vdash \phi$), is a tuple $\Pi = (\mathcal{D}, \mathcal{F})$, where:

    \begin{itemize}
        \item $\mathcal{D} = (V, s, r, p)$ is a derivation tree where $V$ is a finite set of vertices, $s: V \rightarrow Seqs$ is a total function labelling vertices with sequents in $\CLKIDomega$ (resp. $\CLJIDomega$), $r: V \rightarrow Rules$ is a partial function labelling vertices with inference rules in $\CLKIDomega$ (resp. $\CLJIDomega$), $p: \mathbb{N} \times V \rightarrow V$ is a partial function that maps a number $n$ and a vertex $v$, with a vertex $v'$, such that $v'$ is labeled by the $n^{\text{th}}$ premise of the rule $r(v)$ over the sequent $s(v)$, and for every $v \in V$, 
            \begin{prooftree}
                \hypo{s(p_1(v)) \quad \dots \quad  s(p_m(v))}
                \infer1[$(r(v))$]{ s(v)}
            \end{prooftree} 
        is an instance of the rule $r(v)$ in $\CLKIDomega$ (resp. $\CLJIDomega$), with $m$ premises. A vertex $B \in V$ such that $r(B)$ is undefined is called a bud node. The set of bud nodes of a derivation tree $\mathcal{D}$, is written as $Bud(\mathcal{D})$.

        \item $\mathcal{F}: Bud(\mathcal{D}) \rightarrow V \setminus Bud(\mathcal{D})$ is a partial function, that maps each one of the bud nodes in $Bud(\mathcal{D})$ to a non-bud node in the derivation tree, called a companion, satisfying that if $B \in Bud(\mathcal{D})$ and $\mathcal{F}(B) = C$, then 
        $s(B) = s(C)$.
    \end{itemize}

Moreover, the graph of $\Pi$, written as $G_{\Pi}$, is the graph obtained after adding the edges induced by $\mathcal{F}$ and deleting the bud vertices. Specifically, $G_{\Pi} = (V', s, r, p')$ where, $V' = V \setminus Bud(\mathcal{D})$ and $p'$ is defined for any $j$ as: $p'_j(v) = \mathcal{F}(p_j(v))$ if $p_j(v) \in Bud(\mathcal{D})$, or $p'_j(v) = p_j(v)$ otherwise.
\end{definition}

\begin{example}
Figure \ref{exampleCircProof} is an example of a $\CLKIDomega$ cyclic pre-proof, taken from \cite{Brotherston1}. 
We use 
blue/red arrows to connect inductive formulas.
\end{example}

As mentioned before, the Global Trace Condition (GTC) is used as a sufficient criterion to verify the soundness of cyclic pre-proofs in $\CLKIDomega$ and $\CLJIDomega$. To define it, it is first necessary to formalize both the notions of \emph{trace} and \emph{progress point}. 

\begin{definition}[Trace and Progress Point]
    \label{def:traceCyclic}
    Let $\Pi$ be a $\CLKIDomega$ (resp. $\CLJIDomega$) pre-proof and let $(v_i)$ be a sequence of vertices defining a path in $G_{\Pi}$. A trace over $(v_i)$ is a sequence of occurrences $\uptau$ such that, for all $i$:

    \begin{itemize}
          \item $\uptau(i) = (P\mathbf{u},j)$ is in $\Gamma$, where $P$ is an inductive predicate, and $s(v_i) = \Gamma \vdash \Delta$ (resp. $s(v_i) = \Gamma \vdash \phi$).

          \item If $\uptau(i) = (P\mathbf{u},j)$, then $\uptau(i+1) = (Q\mathbf{t},k)$, where $(Q\mathbf{t},k)$ is an occurrence such that $Q$ is an inductive predicate, $(P\mathbf{u},j) \rightarrow_{r(v_i)} (Q\mathbf{t},k)$, and $(Q\mathbf{t},k)$ is in $\Gamma'$, where $s(v_{i+1}) = \Gamma' \vdash \Delta'$ (resp. $s(v_{i+1}) = \Gamma' \vdash \phi'$).

          \item If $r(v_i) = (\text{Case } P)$ is a case split rule, and $\uptau(i) = (P\mathbf{u},j)$ is the progressing occurrence of the rule, then $i$ is said to be a progress point of the trace.
          \item A trace is infinitely progressing if it has infinitely many progress points.
    \end{itemize}  

\end{definition}

\begin{remark}
    \label{rem:traceDef}
    The previous definition of trace corresponds to a proper subset of the original trace relation defined by Brotherston in \cite[Definition 4.2.3]{Brotherston1}. Specifically, our definition of trace corresponds to applying Brotherston's definition to work over occurrences instead of formulas. The reason we use occurrences, is that they allow us to track the essential part of the history of a formula in a pre-proof without ambiguities, and it ensures that two distinct arrows of the trace relation, will never join in the same occurrence of a premise (see Proposition \ref{prop:injective-trace}), a fact which will be exploited in Section \ref{sec:progressDomainCond}.
\end{remark}  

We now define the validity condition for pre-proofs, which we call the global trace condition, following the literature.

\begin{definition}[Global Trace Condition]
    \label{def:GTC}
    A $\CLKIDomega$ (resp. $\CLJIDomega$) pre-proof $\Pi$ is said to be a $\CLKIDomega$ (resp. $\CLJIDomega$) \emph{proof}, if for every infinite path in $G_{\Pi}$, there is an infinitely progressing trace following some tail of the path.
\end{definition}

\begin{example}
    Consider the cyclic pre-proof in Figure \ref{exampleCircProof} and the only infinite path, contained in the cycle between the two right-most branches of the pre-proof. The trace $\uptau = (Ox, Ey, Ey, Ey, Ex, Oy, Oy, Oy)^{\omega}$  is infinitely progressing over this path, since it contains the progressing points $(Ox,Ey)$ and $(Ex,Oy)$. Since this is the only infinite path in the pre-proof, we conclude that the pre-proof satisfies the GTC, and
    that it is a $\CLKIDomega$ proof.
\end{example}

\section{Transition Graphs and Size-Change Termination}
\label{sec:size-change}

Inspired by the work by Ikebuchi \cite{Mirai}, in this section we introduce the notion of {\em transition graphs}, which is intended to characterize the global trace condition from the theory of cyclic proofs in terms of the {\em size-change termination} condition. We consider directed graphs whose edges are labelled with pairs of relations, that we call bi-partite relations. The properties relevant for termination can now be expressed in terms of compositions of such relations.
 
 \subsection{Bi-partite Relations and Transition Graphs}
 
Let $R \subseteq X \times Y$ be a binary relation, then we write $x R y$ for $(x,y) \in R$. 
Thinking of $R$ as a function $R: X \to \wp(Y)$, we set $R(x) = \Set{y \in Y \mid x R y}$. 
Then we define $\dom{R} = \Set{x \in X \mid \exists y \in Y.\; xRy} 
= \Set{x \in X \mid R(x) \neq \emptyset}$, and
$\cod{R} = \Set{y \in Y \mid \exists x \in X.\; xRy} = \bigcup_{x \in X} R(x)$. 

In this work, we shall consider binary relations over some fixed finite set 
$X = \Set{x_1, ..., x_n}$ of abstract ``parameters'', which are consequently themselves finite. We say that $(S,P)$ is a {\em bi-partite} relation over $X$ if both $S, P \subseteq X \times X$ and $S \cap P = \emptyset$, namely if $S, P$ are a partition of the binary relation $S \cup P$ over $X$, with possibly $S$ or $P$ empty.

\begin{definition}[Transition graph]\label{def:transitionGraph}
Let $\biRel$ be a finite set of bi-partite relations over $X$.
A {\em transition graph} of $\biRel$ is a tuple $\Graph = (V, \biRel, E)$ where $V$ is a finite set of vertices, and $E \subseteq V \times \biRel\times V$ is a (finite) set of labeled edges.
\end{definition}

Let $\Graph = (V, \biRel, E)$; we abbreviate $(u, (S,P), v) \in E$ by $\edgeG{(S,P)}{u}{v}{\Graph}$.
Also, if $e = \edgeG{(S,P)}{u}{v}{\Graph}$ is any edge, 
then $\source{e} = u$, $\target{e} = v$, and $\lbl{e} = (S,P)$ are the {\em source}, {\em target}, and {\em label}
of $e$ respectively. When $\Graph$ is understood, we omit the subscripts in $\edge{(S,P)}{u}{v}$. 

Suppose that $\edgeG{(S,P)}{u}{v}{\Graph}$, and $x, y \in X$. When $x S y$ we say that the size associated with $x$ in the abstract state $u$ is equal or greater than the size associated with $y$ in $v$ in some intended ordering; hence, if $x S y$ we say that 
$x,y$ are {\em stationary} in the transition from $u$ to $v$. If instead  $x P y$, then the size of $x$ in $u$ is strictly greater than that of $y$ in $v$, and 
we say that $x$ {\em progresses} to $y$ in the transition from $u$ to $v$. 
Finally, if $x$ and $y$ are unrelated by $S \cup P$, this
represents the lack of knowledge about the ordering of their sizes in $u$ and $v$, respectively.

\medskip
Transition graphs are a variant of the notion of {\em size-change graph} as defined in \cite[Def. 2.1]{Lee3}, abstractly representing the labeled transition system of the flow of values of certain variables through the calls of functions in a functional program, where labels convey the information needed to detect termination.
More precisely, a size-change graph $G:\texttt{f} \to \texttt{g}$ for a pair of function names $\texttt{f}$, 
$\texttt{g}$ out of some functional program $\texttt{p}$
is given by a triple $G = (\textit{Param}(\texttt{f}), \textit{Param}(\texttt{g}), \textsf{E})$, where $\textit{Param}(\texttt{f}), \textit{Param}(\texttt{g})$
are disjoint sets of variables occurring in the definitions of $\texttt{f}$ and $\texttt{g}$, and $\textsf{E} \subseteq \textit{Param}(\texttt{f}) \times \Set{\leqdownarrow, \downarrow} \times \textit{Param}(\texttt{g})$ and there are not two edges from $x \in \textit{Param}(\texttt{f})$ to
$y \in \textit{Param}(\texttt{g})$ labelled with both $\leqdownarrow$ and
$\downarrow$ in $\textsf{E}$.

In our view, $G:\texttt{f} \to \texttt{g}$, 
$G = (\textit{Param}(\texttt{f}), \textit{Param}(\texttt{g}), \textsf{E})$ corresponds 
to the edge $\edgeG{(S_G,P_G)}{f}{g}{\Graph}$ of a transition graph $\Graph$, and
$X$ is the set of all parameters occurring in the program $\texttt{p}$, 
and $S_G = \Set{(x,y) \mid (x, \leqdownarrow, y) \in \textsf{E}}$ and 
$P_G = \Set{(x,y) \mid (x, \downarrow, y) \in \textsf{E}}$ are binary relations over $X$.
The transition graph $\Graph$ has vertices $V=$ the set of all functions $\texttt{f}$
defined in the program $\texttt{p}$, and has set of
bi-partite relations $\biRel = \Set{(S_G,P_G) | G \mbox{ size-change graph}}$.

A {\em multipath} \cite[Def. 2.3]{Lee3} is the composition of finitely or infinitely many size-change graphs. 
The composition $GG'$ of $G:\texttt{f} \to \texttt{g}$ with $G':\texttt{g} \to \texttt{h}$
requires that the target function symbols of $G$ and the source function symbol of $G'$ coincide. 
In the case of our transition graphs, we have a unique graph $\Graph$, with finite set of parameters $X \supseteq
 \textit{Param}(\texttt{f}) \cup \textit{Param}(\texttt{g}) \cup \textit{Param}(\texttt{h})$,
and we express the composition $GG'$ by composing 
the edges $\edgeG{(S_G,P_G)}{\texttt{f}}{\texttt{g}}{\Graph}$ and  
$\edgeG{(S_{G'},P_{G'})}{\texttt{g}}{\texttt{h}}{\Graph}$.

\subsection{Composition of bi-partite Graphs and Size Change Termination}

Binary relations compose: if $R \subseteq X \times Y$ and 
$R' \subseteq Y \times Z$ then $RR' \subseteq X \times Z$ is such that
$xRR'z$ if and only if there is $y \in Y$ s.t. $xRy$ and $yR'z$, a fact 
we abbreviate by $xRyR'z$. Relation compositions are associative 
and have the identities $I_X = \Set{(x,x) \mid x \in X}$ as units. We can extend composition
to bi-partite relations in a way compatible with the information weakly decreasing/strictly
decreasing associated to the partition. 
In the next definition, we say that $x$ progresses to $z$ in this composition if for some $y$ either $x$ progresses to $y$ such that $y(S' \cup P')z$ or if $x,y$ are stationary and $yP'z$. We say that $x,z$ are stationary in the composition $(S, P) \comp (S', P')$ if there is a $y$ such that $x$ is stationary to $y$ which is stationary to $z$, provided $x$ \emph{does not progress} to $z$. In
this case we consider the progression information to be more relevant and we drop
the stationary information from $(S, P) \comp (S', P')$.

\begin{definition}[Composition of bi-partite relations]\label{def:bi-partiteRelations}
Let $(S,P)$, $(S',P')$ be bipartite relations over $X$; then we define their {\em composition} $(S, P) \comp (S', P') = (S'' \setminus P'',\; P'')$
where
\[S'' = S S' \qquad \mbox{and} \qquad P'' = (S P') \cup (P S') \cup (P P') \]
We call $|(S,P)| = S \cup P$ the relation associated to $(S,P)$.
\end{definition}

We will prove that the case all relations are injective the definition
can be replaced with $(S, P) \comp (S', P') = (S'', \; P'')$
The composition of bi-partite relations is bi-partite by construction; also, it is associative
with identity the pair $(I_X, \emptyset)$.
We observe that composition is compatible with the operation 
$|\cdot|$ and ordinary composition of binary relations: 
\begin{align}
\label{eq:unionComp}
|(S,P) \comp (S',P')| = |(S'' \setminus P'', P'')| &= (S''\setminus P'') \cup P''  
= S'' \cup P''  \nonumber \\ = SS' \cup  (S P') \cup (P S') \cup (P P')
& = (S \cup P) (S' \cup P') = |(S,P)| |(S',P')|
\end{align}
Let $\biRel^*$ be the closure under composition $\comp$ of a set $\biRel$ of bi-partite relations. Notice that the set of bi-partite relations over the finite set $X$
is a subset of $\wp(X^2) \times \wp(X^2)$ which is finite, hence $\biRel^*$ is finite.
The role of a transition graph $\Graph$ with labels in $\biRel$ is to reduce the number of
 compositions of bipartite relations in $\biRel^*$ which are accepted. We formalize this notion
by adapting the usual definition of transitive closure of a graph.

\begin{definition}[Transitive closure of a transition graph]\label{def:transitiveClosure}
We say that  the graph $\Graph^* = (V, \biRel^*, E^*)$ is the {\em transitive closure} of $\Graph = (V, \biRel, E)$ if
$E^* \subseteq V \times \biRel^* \times V$ is the smallest set such that $E \subseteq E^*$ and
\[ (u, (S,P), v), (v, (S',P'), w) \in E^* \Then (u, (S,P) \comp (S',P'), w) \in E^*\]
If $(u, (S,P) \comp (S',P'), w) \in E^*$ for some vertexes $u, w$ then we say that $(S,P)$ and $(S',P')$ are {\em composable} in $\Graph^*$; we say
that $(S,P)$ is {\em self-composable} in $\Graph^*$ if $(v, (S,P), v) \in E^*$ for some vertex $v$.
\end{definition}

In words, $\Graph^*$ is the transitive closure of the graph $\Graph$ with edges labeled by the composition of the labels of composable edges in $\Graph$. 
We could say that it is essentially a category, but for the lack of identities.

Let $\Graph^* = (V, \biRel^*, E^*)$; we abbreviate $(u, (S,P), v) \in E^*$ by 
$\edgeGstar{(S,P)}{u}{v}{\Graph}$. 
If $\edgeGstar{(S,P)}{v}{v}{\Graph}$ is self-composable then $\edgeGstar{(S,P)^n}{v}{v}{\Graph}$ for all $n > 0$ by definition of composition over the edges 
of $\Graph^*$, where $(S,P)^n$ is the composition of $(S,P)$ with itself $n$-times.
As before, we omit the subscripts in $\edgestar{(S,P)}{u}{v}$, when $\Graph$ is understood.
In order to state GTC for transition graphs we need the notions of sequence and suffix.

\begin{definition}[Sequence and Inhabited Suffix]
Let $\alpha \leq \omega$ be an ordinal, 
identified with an initial segment of $\Nat$ 
ordered in the natural way, and possibly coinciding with $\Nat$. Let $k \in \alpha$.
\begin{enumerate}
\item
a sequence over some set $A$ with domain $\alpha$ is just a map $s:\alpha \to A$;  

\item
$\alpha\uparrow_k \;=\; \Set{i \in \alpha \mid i \geq k}$ is an inhabited suffix of $\alpha$;

\end{enumerate}
\end{definition}

We can now restate the notion of trace and the Global Trace Condition for transition graphs.

\begin{definition}[Trace, path and progress]\label{def:TracePath}
Let $\Graph = (V, \biRel, E)$ be a transition graph and $\alpha \leq \omega$, $k \in \alpha$; 
then:
\begin{enumerate}[i)]
\item 
a {\em trace} is a map $\tau:\alpha\uparrow_k \to X$ with domain some inhabited 
suffix of $\alpha$;

\item 
a {\em path} is a map $\ppath:\alpha \to E$ such that if $i + 1 \in \alpha$ then 
$\target{\ppath(i)} = \source{\ppath(i+1)}$;

\item 
a trace $\tau:\alpha\uparrow_k \to X$ {\em follows} a path $\ppath: \alpha \to E$,
written $\tau \models \ppath$, if for all $i,i+1 \in \alpha\uparrow_k$,
if $\lbl{\ppath(i)} = (S,P)$ then $\tau(i) (S \cup P) \tau(i + 1)$.

If furthermore $\tau(i) P \tau(i + 1)$ 
we say that $i$ is a {\em progress point} of $\tau$ in $\ppath$.

\item 
a trace $\tau$ such that $\tau\models \ppath$ 
is infinitely progressing if $\tau$ has infinitely many progress points in $\ppath$.
\end{enumerate}
\end{definition}

\begin{remark}
A trace $\tau = x_k, x_{k+1}, x_{k+2} \cdots$ follows a path $\ppath: v_0 \stackrel{(S_0,P_0)}{\longrightarrow} v_1 \stackrel{(S_1,P_1)}{\longrightarrow} v_2 \cdots$
if for $\tau$, $\ppath$ have domain $\alpha\uparrow_k$, $\alpha$, respectively, 
and $x_i (S_i \cup P_i) x_{i+1}$ for all $i, i+1 \in \alpha\uparrow_k$. 
Hence, for $\tau\models\ppath$ to hold, it suffices that the above relation holds, possibly disregarding an initial segment of $\ppath$ which is a proper subset.
In the particular case that $\alpha < \omega$, namely when both $\tau$ and $\ppath$ are finite, there is no condition about $x_{\alpha - 1}$ w.r.t. $S_{\alpha - 1} \cup P_{\alpha - 1}$, but this is immaterial for our development, 
as for us the only interesting paths and traces are the infinite ones.
\end{remark}

We say that $\tau$ is a trace of $\Graph$ if 
$\tau \models \ppath$ for some path $\ppath$ of $\Graph$.
In our terminology, the path $\ppath$ and the trace $\tau$ correspond to a {\em multipath} $\mathcal{M}$ and a {\em thread} in \cite{Lee1}, respectively.
Both $\Graph$ and $\Graph^*$ are finite, so that an infinite path $\ppath$ exists in $\Graph$ only if it has a cycle, and therefore there exists
some $(S,P) \in \biRel^*$ which is self-composable in $\Graph^*$.

We are eventually in place to define the size-change termination condition for transition graphs, which is essentially the same as \cite[Def. 2.6]{Lee3}.

\begin{definition}[Size-change Termination - SCT]
\label{def:SCT}
The transition graph $\Graph$ satisfies the {\em Size-change Termination} condition 
if for all infinite path $\ppath$ of $\Graph$ there is some infinitely progressing trace $\tau$ 
such that $\tau \models \ppath$.
\end{definition}

In the rest of this section, we relate the progress of a trace $\tau$ w.r.t. $\ppath$ 
to the composition $(S,P)$ of the bi-partite relations labeling the edges in $\ppath$,
and show that progressing is equivalent to having $x, y$ in the image of $\tau$ such that $x$ precedes $y$ in $\tau$ and $x P y$.

\begin{lemma}\label{lem:P-lemma}
Let $(S,P) = (S_1,P_1)\comp (S_2,P_2)$ and $x (S_1 \cup P_1) y (S_2 \cup P_2) z$, then
\begin{enumerate}[i)]

\item
\label{lem:P-lemma-1} $x P_1 y \Then x P z$

\item
\label{lem:P-lemma-2} $y P_2 z \Then x P z$

\end{enumerate}
\end{lemma}

If $\ppath: v_0 \stackrel{(S_0,P_0)}{\longrightarrow} v_1 \cdots \; v_{b-1} \stackrel{(S_{b-1},P_{b-1})}{\longrightarrow} v_{b}$ is a (finite) path of $\Graph$, for $0 \leq i < j < b$
we abbreviate $(S_{i,j}, P_{i,j}) = \bigcomp_{p = i}^{j} (S_p,P_p) = (S_i,P_i) \comp \cdots \comp (S_j,P_j)$.

\begin{proposition}\label{prop:progresing-and-P}
Let $\ppath: v_0 \stackrel{(S_0,P_0)}{\longrightarrow} v_1 \cdots \; v_{b-1} \stackrel{(S_{b-1},P_{b-1})}{\longrightarrow} v_{b}$ be a finite path of $\Graph$
and $\tau = x_a, \ldots , x_{b-1}$ be a trace such that $\tau \models \ppath$. Then $\tau$ has a progress point in $\ppath$ if and only if
there exists $a \leq h < b$ such that
\[x_h P_{h,b-1} x_b \And \forall h \leq j < b.\; x_j (S_j \cup P_j) x_{j+1}\]
\end{proposition}

\section{Transition Graphs of Cyclic Proofs}
\label{sec:transition-graph-proofs}

Given a $\CLKIDomega$ or $\CLJIDomega$ pre-proof $\Pi$, we show how to construct the induced transition graph with bi-partite relations. Our approach follows Ikebuchi's work \cite{Mirai}, where the equivalence between the GTC and SCT is formally stated by extracting the size-change graphs of a cyclic proof. In our case, we extract a transition graph with bi-partite relations of a cyclic proof, which in principle corresponds to the same process, with some simplification in the case of an injective trace relation.

\begin{definition}[Transition Graph of a Pre-Proof]
\label{def:preProofTrans}

Let $\Pi = ((V, s, r, p), \mathcal{F})$ be a $\CLKIDomega$ pre-proof with a graph $G_{\Pi} = (V',s,r,p')$. We note $\mathcal{G}_{\Pi} = (V', \mathcal{R}, E)$ as the transition graph over bi-partite relations induced by $\Pi$, where:
\begin{itemize}

     \item $\mathcal{R} \subseteq X \times X$ where $X = \{ (P\mathbf{u},j) \ | \ \text{$P$ is inductive, $\exists v \in V'$ s.t. $s(v) = \Gamma \vdash \Delta$ and $(P\mathbf{u},j) \in \Gamma$} \}$

     \item Let $v,v' \in V'$ be two vertices which share an edge, i.e., there exists a $j$ such that $p'_j(v) = v'$. Moreover, let $s(v) = \Gamma \vdash \Delta$ (resp. $s(v') = \Gamma' \vdash \Delta'$) and let $\Psi \subseteq \Gamma$ (resp. $\Theta \subseteq \Gamma'$)
     be the sub-sequences containing only the occurrences of $\Gamma$ (resp. $\Gamma'$), of the form $(P\mathbf{u}, i)$, where $P$ is an inductive predicate. We define the bi-partite relation $(S,P)_{v,v'}$ as follows:
        \begin{enumerate}[i)]
            \item The domain and codomain of $|(S,P)_{v,v'}|$ are $\Psi$ and $\Theta$ respectively.

            \item If $r(v)$ is not a case split rule, then for all $(P\mathbf{u},i)$ in $\Psi$ and $(Q\mathbf{t},j)$ in $\Theta$, $(P\mathbf{u},i) S (Q\mathbf{t},j)$ if and only if $(P\mathbf{u},i) \rightarrow_{r(v)} (Q\mathbf{t},j)$.

            \item If $r(v)$ is a case split rule, then for all $(Q\mathbf{t},j)$ in $\Theta$, $(P\mathbf{u},i) P (Q\mathbf{t},j)$ if and only if $(P\mathbf{u},i) \rightarrow_{r(v)} (Q\mathbf{t},j)$ and $(P\mathbf{u},i)$ is the progressing 
            occurrence in $\Gamma$. Otherwise, if $(P\mathbf{u},i)$ is stationary, then $(P\mathbf{u},i) S (Q\mathbf{t},j)$.

        \end{enumerate}

    \item $\mathcal{R} = \{(S,P)_{v,v'} \ | \ \text{for all $v,v' \in V$ s.t. $p'_j(v) = v'$ for some $j$}  \}$

    \item $E = \{\edge{(S,P)_{v,v'}}{v}{v'} \ | \ \text{for all $v,v' \in V$ s.t. $p'_j(v) = v'$ for some $j$}  \}$
 \end{itemize} 

This process can also be applied to $\CLJIDomega$ pre-proofs, using sequents of the form $\Gamma \vdash \phi$.

\end{definition}

Note that this construction defines an isomorphism between graphs. Specifically, $G_{\Pi}$ is isomorphic to $\mathcal{G}_{\Pi}$, in the sense that for all $v,v' \in V'$ then $p'_j(v) = v'$ in $G_{\Pi}$ if and only if $\edge{(S,P)_{v,v'}}{v}{v'}$ in $\mathcal{G}_{\Pi}$.

\begin{figure}[t]

\begin{mdframed}
\centering
\resizebox{0.60\textwidth}{!}{%
\begin{circuitikz}
\tikzstyle{every node}=[font=\Large]

\draw[rounded corners]  (2.5,13) rectangle  node {\LARGE $Ex$} (3.75,11.75);
\draw[rounded corners]  (6.25,13) rectangle  node {\LARGE $Oy$} (7.5,11.75);
\draw[rounded corners]  (10,13) rectangle  node {\LARGE $Oy$} (11.25,11.75);
\draw[rounded corners]  (2.5,9.25) rectangle  node {\LARGE $Ox$} (3.75,8);
\draw[rounded corners]  (6.25,9.25) rectangle  node {\LARGE $Ey$} (7.5,8);
\draw[rounded corners]  (10,9.25) rectangle  node {\LARGE $Ey$} (11.25,8);
\draw[rounded corners]  (-1.25,11.25) rectangle  node {\LARGE $\emptyset$} (0,10);
\draw[rounded corners]  (6.25,16) rectangle  node {\LARGE $\emptyset$} (7.5,14.75);
\draw[rounded corners]  (10,16) rectangle  node {\LARGE $\emptyset$} (11.25,14.75);

\node [font=\Large] at (3.25,11.25) {$v_3$};
\node [font=\Large] at (7,11.25) {$v_4$};
\node [font=\Large] at (10.75,11.25) {$v_5$};
\node [font=\Large] at (3.25,7.5) {$v_6$};
\node [font=\Large] at (7,7.5) {$v_7$};
\node [font=\Large] at (10.75,7.5) {$v_8$};
\node [font=\Large] at (-0.5,9.5) {$v_0$};
\node [font=\Large] at (7,14.25) {$v_1$};
\node [font=\Large] at (10.75,14.25) {$v_2$};

\draw [ color={rgb,255:red,204; green,0; blue,0}, line width=1pt, ->, >=Stealth] (3.75,12.25) -- (6.25,12.25)node[pos=0.5, fill=white]{\normalsize $(Ex,Oy)$};
\draw [ color={rgb,255:red,26; green,95; blue,180}, line width=1pt, ->, >=Stealth] (7.5,12.25) -- (10,12.25)node[pos=0.5, fill=white]{\normalsize $(Oy,Oy)$};
\draw [ color={rgb,255:red,204; green,0; blue,0}, line width=1pt, ->, >=Stealth] (3.75,8.5) -- (6.25,8.5)node[pos=0.5, fill=white]{\normalsize $(Ox,Ey)$};
\draw [ color={rgb,255:red,26; green,95; blue,180}, line width=1pt, ->, >=Stealth] (7.5,8.5) -- (10,8.5)node[pos=0.5, fill=white]{\normalsize $(Ey,Ey)$};
\draw [ color={rgb,255:red,26; green,95; blue,180}, line width=1pt, ->, >=Stealth] (10.5,11.75) -- (3.75,9.25)node[pos=0.8, fill=white]{\normalsize $(Oy,Ox)$};
\draw [ color={rgb,255:red,26; green,95; blue,180}, line width=1pt, ->, >=Stealth] (10.5,9.25) -- (3.75,11.75)node[pos=0.2, fill=white]{\normalsize $(Ey,Ex)$};

\draw [line width=1pt, ->, >=Stealth] (0,10.5) -- (2.5,12.25)node[pos=0.5, fill=white]{$\emptyset$};
\draw [line width=1pt, ->, >=Stealth] (0,10.5) -- (2.5,8.5)node[pos=0.5, fill=white]{$\emptyset$};
\draw [line width=1pt, ->, >=Stealth] (7.5,15.5) -- (10,15.5)node[pos=0.5, fill=white]{$\emptyset$};
\draw [line width=1pt, ->, >=Stealth] (3,13) -- (6.25,15.25)node[pos=0.5, fill=white]{$\emptyset$};
\end{circuitikz}
}%

\end{mdframed}

\caption{The transition graph of the cyclic pre-proof in Figure \ref{exampleCircProof}, where the set of parameters is the set $X = \{ (Ex,0), (Ey,0), (Ey,1), (Ox,0), (Oy,0), (Oy,1)\}$ of occurrences of the inductive predicates in the pre-proof. 
For each bi-partite relation
$(S,P)_{v,v'}$, the relation $S$ is represented by {\color{blue} \bf blue} pairs, while the pairs in $P$ are colored in {\color{red} \bf red}. In this case, there is only one pair in each bi-partite relation $(S,P)_{v,v'}$, therefore, we label the edges with just one pair to simplify the presentation.}

\label{exampleTransGraph}
\end{figure}
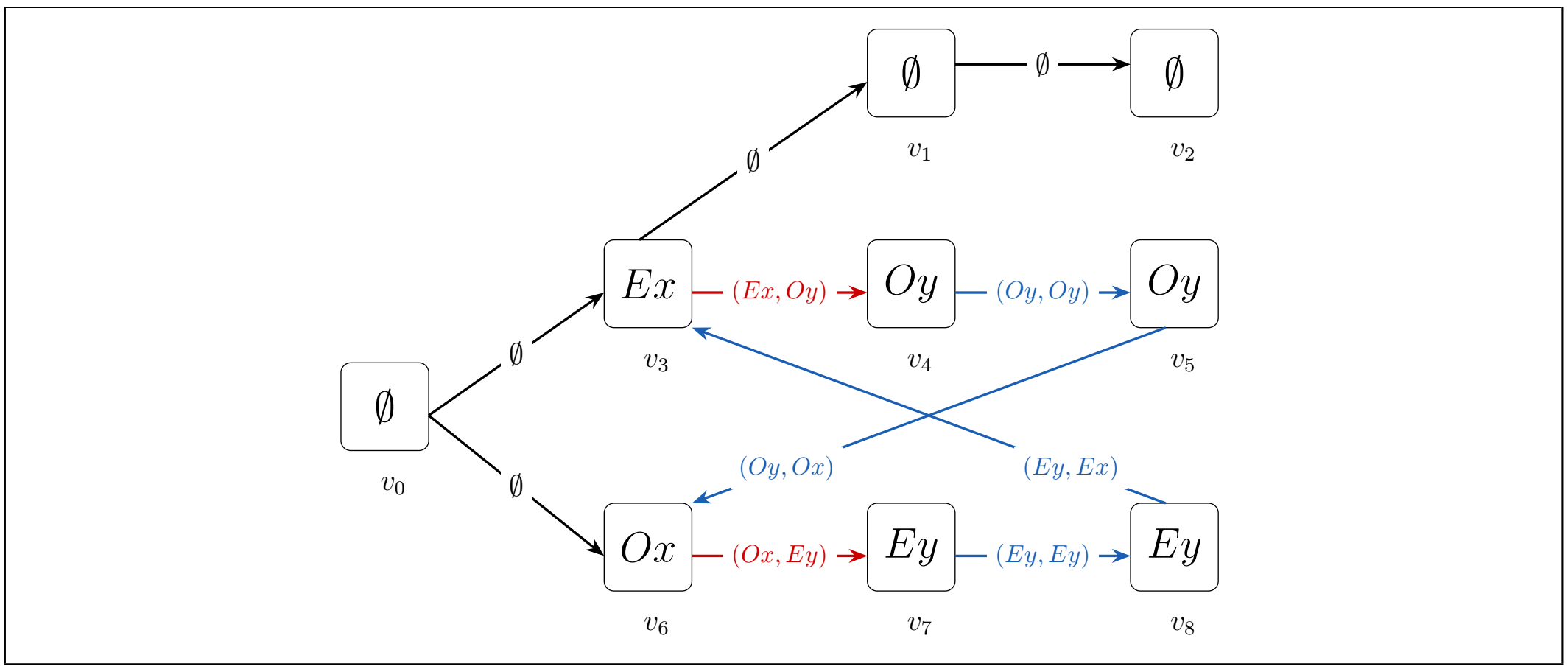

\begin{example}
    Using Definition \ref{def:preProofTrans} over the cyclic pre-proof in Figure \ref{exampleCircProof}, we obtain the transition graph in Figure \ref{exampleTransGraph}. 
\end{example}

Using the construction specified in Definition \ref{def:preProofTrans}, and the Definitions \ref{def:TracePath}, inspired by SCT, and \ref{def:traceCyclic} of path and traces for transition graphs and cyclic pre-proofs, it can be deduced that they are the same:

\begin{proposition}
Consider a circular pre-proof $\Pi$ with a graph $G_{\Pi}$ and induced transition graph $\mathcal{G}_{\Pi}$. Moreover, let $\pi$ be a sequence of vertices defining a path in $G_{\Pi}$, and $\pi'$ corresponding to the same path in $\mathcal{G}_{\Pi}$, but defined as in Definition \ref{def:TracePath} (we can do this, since the two graphs are isomorphic). Then, $\uptau$ is a trace over $\pi$ if and only if $\uptau \models \pi'$.
\end{proposition}

Then, using the previous proposition and the Definitions \ref{def:GTC} and \ref{def:SCT}, the equivalence between GTC and SCT follows:

\begin{proposition}\label{prop:GTC-SCT}
Let $\Pi$ be a $\CLKIDomega$ (resp. $\CLJIDomega$) pre-proof. Then $\Pi$ is a $\CLKIDomega$ (resp. $\CLJIDomega$) proof if and only if $\mathcal{G}_{\Pi}$ satisfies the size-change termination condition.
\end{proposition}

\section{Injectivity and Progress Domain Condition}
\label{sec:progressDomainCond}

We define injectivity over relations and prove some of its properties. Our goal is finding a case in which GTC is simpler to state. A relation $R \subseteq X \times Y$ is {\em injective} if for all $x,x' \in X$ and $y \in Y$, $xRy$ and $x'Ry$ implies $x = x'$. Equivalently, $R$ is injective if all the $R(x)$ for $x \in X$ are disjoint. Moreover, the composition of two injective relations is always injective.

\begin{lemma}\label{lem:comInjective}
If both $R \subseteq X \times Y$ and $R' \subseteq Y \times Z$ are injective, then $RR' \subseteq X \times Z$ is injective.
\end{lemma}

\begin{proof}
If $xRR'z$ and $x'RR'z$ then there are $y,y' \in Y$ such that $xRy$ and $yR'z$, and $x'Ry'$ and $y'R'z$. By the injectivity of $R'$ we have $y = y'$,
hence $x = x'$ by the injectivity of $R$.
\end{proof}

We say that a bi-partite relation $(S, P)$ is {\em injective} if $|(S, P)| = S \cup P$ is such in the above sense. We can then extend the previous lemma in the case of injective bi-partite relations.

\begin{lemma}\label{lem:biRelInjectiveComp}
\begin{enumerate}
\item
If both $(S,P)$ and $(S',P')$ are injective, then $(S,P) \comp (S',P')$ is injective.

\item
If $(S',P')$ is injective, then composition can be simplified by:
\[ (S,P) \comp (S',P') = (SS', SP' \cup PS' \cup PP') \]
\end{enumerate}
\end{lemma}

\begin{proof}
\begin{enumerate}
\item
By (\ref{eq:unionComp}) in Section \ref{sec:size-change},  we know that $|(S,P) \comp (S',P')| =  |(S,P)| |(S',P')|$ which is ordinary composition of the injective relations $S \cup P$ and $S' \cup P'$, hence it is injective by Lemma \ref{lem:comInjective}.

\item
By definition the stationary component of $(S,P) \comp (S',P')$ is $S'' \setminus P''$
where $S'' =S S'$ and $P'' = SS', SP' \cup PS' \cup PP'$. Our thesis becomes:
$S'' \setminus P'' = S S'$. We have to prove that $S''$ and $P''$ are disjoint. 
We assume $xSyS'z$ and $x (S \cup P) y' (S \cup P) z$ 
(there is some other trace from $x$ to $y$) 
in order to prove that this trace is not in $P''$, namely that
$\neg (x P  y')$ and $\neg (y' P  z)$.  From $S' \cup P'$ injective and $y (S' \cup P') z$
we deduce that $y = y'$. Our thesis is now $\neg (x P  y)$ and $\neg (y P  z)$.
This follows from $xSy$ and $yS'z$ and $S \cap P = S' \cap P' = \emptyset$.
\end{enumerate}
\end{proof}

We say that the set of bi-partite relations $\biRel$ is {\em injective} if all $(S,P) \in \biRel$ are such. If $\Graph = (V, \biRel, E)$ with injective $\biRel$, by abuse of language we say that $\Graph$ is injective. By Lemma \ref{lem:biRelInjectiveComp}, if $\biRel$ is injective, then its composition closure $\biRel^*$ is such. 

The notion of injectivity is crucial when analyzing the structure of transition graphs arising from pre-proofs in $\CLJIDomega$ and $\CLKIDomega$. In particular, the construction of the graph $\mathcal{G}_{\Pi}$ associated with a pre-proof $\Pi$ with contexts as lists, as defined in the previous sections, is always injective. 
This observation leads to the following result:

\begin{theorem}\label{thr:injective-pre-proofs}
Let $\Pi$ be any $\CLJIDomega$ or $\CLKIDomega$ pre-proof. 
Then $\Graph_\Pi$ is injective.
\end{theorem}

\begin{proof}
By Prop. \ref{prop:injective-trace}.\ref{prop:injective-trace-02}.
\end{proof}

\begin{remark}
The original version of traces by Brotherston and Simpson, in the system named $\CLKIDomega$, 
is \emph{not} fan-in free. As a counterexample, consider the following instance of left-equality 
rule, using sequents as sets, with $N$ the inductive predicate of Example \ref{ex:NEO}.

 \begin{center}
    \begin{prooftree}
        \hypo{ \Gamma, \vnode1{Nt} \vdash \Delta}
        \infer1[(=L)]{\Gamma, t = u, \vnode2{Nt}, \vnode3{Nu} \vdash \Delta} 
    \end{prooftree}

    $
    \dirflowblueedges{node3/node1, node2/node1} 
    $
  \end{center}
  
If we apply Brotherston's definition of trace in
\cite[Definition 4.2.3]{Brotherston1}, we obtain that the inductive formula $Nt$ in the premise has at least two incoming edges obtained from the trace relation, from the inductive formulas $Nt$ and $Nu$ in the conclusion. Even considering sequents as multisets, fan-in free connections between inductive formulas are not guaranteed. For example, consider the left-weakening rule:
    
    \begin{center}
      \begin{prooftree}
        \hypo{\Gamma, \vnode1{Nu} \vdash \Delta}
        \infer1[(WeakL)]{\Gamma, \vnode2{Nu}, \vnode3{Nu} \vdash \Delta} 
      \end{prooftree}

    $
    \dirflowblueedges{node3/node1, node2/node1} 
    $  
    \end{center}

\noindent
where one of the inductive predicates $Nu$ in the conclusion is a direct descendant of the inductive predicate in the premise, while the other was freshly introduced by the weakening rule. Again, we apply Brotherston's definition of trace then the last two 
occurrences of $Nu$
in the right-hand side of conclusion are connected to the same 
occurrence of $Nu$ in the assumption
of the rule.
\end{remark}

In the previous section, we have seen that the SCT condition for transition graphs and the GTC condition for cyclic proofs coincide up to rephrasing GTC in terms of transition graphs. 
We have also proved that the transition graphs associated with cyclic pre-proofs in systems $\CLJIDomega$ and $\CLKIDomega$ are injective. In this section, we formalize a new condition
on transition graphs, which we call {\em Progress Domain Condition}, shortly PDC, and show that, although it is strictly weaker than SCT, it implies SCT in the case of injective graphs. Hence PDC is a simpler reformulation of GTC for our deduction system.

\begin{definition}[Progress Domain Condition - PDC]\label{def:PDC}
A transition graph $\Graph = (V,\biRel,E)$ 
satisfies the {\em Progress Domain Condition} 
if for all self-composable $(S,P) \in \biRel^*$ there are $x, y \in  \dom{S \cup P}$ such that
$x$ progresses to $y$. In other words we require:
\[\forall (S,P) \in \biRel^*, v \in V.\; (\edgestar{(S,P)}{v}{v}) \Then \exists x, y \in \dom{S \cup P}.\; x P y \]
\end{definition}

\begin{proposition}\label{prop:GTCimplPDC}
If $\Graph$ satisfies SCT, then it satisfies PDC.
\end{proposition}

\begin{proof}
Let $\edgestar{(S,P)}{v}{v}$ be a self-composable path of $\Graph^*$. 
Then there is an infinite path
\[\ppath : v \stackrel{(S,P)}{\longrightarrow} v \stackrel{(S,P)}{\longrightarrow} v \stackrel{(S,P)}{\longrightarrow}\cdots\]
By SCT, there is a trace $\tau = x_k, x_{k+1}, \ldots$ such that $\tau\models\ppath$ 
and $\tau$ is infinitely progressing
$\tau$ must progress in some step $\edgestar{(S,P)}{v}{v}$.

By Proposition \ref{prop:progresing-and-P}, it follows that $x_a P x_{a+1}$ for some 
$a \ge k$; now $x_a, x_{a+1} \in \dom{S \cup P}$ since
$\tau\models \ppath$. It follows that $\Graph$ satisfies PDC.
\end{proof}

\begin{remark}\label{rem:PDCnotGTC}
The inverse of Proposition \ref{prop:GTCimplPDC} does not hold without injectivity. 
Consider $X_0 = \Set{x_1, x_2}$ and $\biRel_0 = \Set{(S_0,P_0)}$ where
$S_0 = \Set{(x_1, x_1)}$ and $P_0 = \Set{(x_2, x_1)}$. Let $\Graph_0 = (\Set{v}, \biRel_0, \Set{\edge{(S_0,P_0)}{v}{v}})$. Then $(S_0,P_0)$
is self-composing, $x_1, x_2 \in \dom{S_0 \cup P_0}$ and $x_2 P_0 x_1$, hence $\Graph_0$ satisfies the PDC.

On the other hand, the infinite path $\ppath_0$ represented in Figure \ref{fig:PDC-counterexample} is such that the only infinite traces that follow it are
$\tau_0 : x_1 , x_1, x_1, x_1, \cdots$ and $\tau_1 : x_2, x_1 , x_1, x_1, \cdots$, 
and neither of them is infinitely progressing. 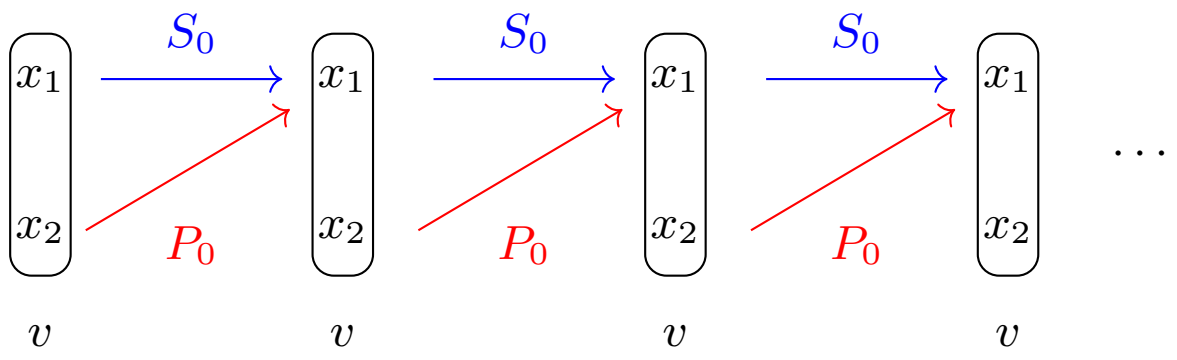
\begin{figure}[h]
\vspace{0.5cm}
\centering
\begin{tikzpicture}[scale=1, every node/.style={font=\small}]

\def\xstep{2.2}

\begin{scope}[xshift=0*\xstep cm]
  
  \draw[black, rounded corners] (-1.2,0.8) rectangle (-0.8,-0.8);
  \draw[black, rounded corners] (0.8,0.8) rectangle (1.2,-0.8);

  \node at (-1, 0.5) {$x_1$};
  \node at (1, 0.5) {$x_1$};
  \node at (-1,-0.5) {$x_2$};
  \node at (1,-0.5) {$x_2$};

  \node at (-1, -1.2) {$v$};
  \node at (1, -1.2) {$v$};
  
  \draw[->, blue] (-0.6, 0.5) -- (0.6, 0.5);

  \draw[->, red] (-0.7, -0.5) -- (0.65, 0.30);

  \node at (0, 0.8) {\textcolor{blue}{$S_0$}};

  \node at (0, -0.6) {\textcolor{red}{$P_0$}};
  
\end{scope}

\begin{scope}[xshift=1*\xstep cm]
  \node at (1, 0.5) {$x_1$};
  \node at (1,-0.5) {$x_2$};

  \draw[black, rounded corners] (0.8,0.8) rectangle (1.2,-0.8);

  \node at (1, -1.2) {$v$};
  
  \draw[->, blue] (-0.6, 0.5) -- (0.6, 0.5);

  \draw[->, red] (-0.7, -0.5) -- (0.65, 0.30);

  \node at (0, 0.8) {\textcolor{blue}{$S_0$}};

  \node at (0, -0.6) {\textcolor{red}{$P_0$}};
\end{scope}

\begin{scope}[xshift=2*\xstep cm]
  \node at (1, 0.5) {$x_1$};
  \node at (1,-0.5) {$x_2$};

  \draw[black, rounded corners] (0.8,0.8) rectangle (1.2,-0.8);

  \node at (1, -1.2) {$v$};
  
  \draw[->, blue] (-0.6, 0.5) -- (0.6, 0.5);

  \draw[->, red] (-0.7, -0.5) -- (0.65, 0.30);

  \node at (0, 0.8) {\textcolor{blue}{$S_0$}};

  \node at (0, -0.6) {\textcolor{red}{$P_0$}};
\end{scope}

\begin{scope}[xshift=3*\xstep cm]
  \node at (-0.3, 0.0) {$\cdots$};
\end{scope}

\end{tikzpicture}
\caption{Infinite path $\ppath_0 : v \stackrel{(S_0,P_0)}{\longrightarrow} v \stackrel{(S_0,P_0)}{\longrightarrow} v \;\cdots$ in $\mathcal{G}_0$}
\label{fig:PDC-counterexample}
\end{figure}

\end{remark}

The key point in the counterexample of Remark \ref{rem:PDCnotGTC} is that $\biRel_0$ is not injective, both $x_1$ and $x_2$ are related with $x_1$. 
Toward establishing that PDC implies SCT
in case of an injective $\biRel$, we first prove the lemma below for transitive relations in 
$\biRel$. 

We say that $(S,P)$ is {\em transitive} if $S \cup P$ is such, namely $x(S \cup P)y(S \cup P)z$ implies $x(S \cup P) z$, for all $x, y, z \in X$. We first prove that PDC implies the so-called
\quotationMarks{algorithmic criterion for SCT} (\cite{Lee1},Section 3.2).

\begin{lemma}
\label{lem:transitivity}
Suppose that $\Graph$ is injective and satisfies PDC, and that 
$(S,P)$ is self-composing in $\Graph$ 
\begin{enumerate}
\item
\label{lem:transitivity-01}
If $(S,P)$ is transitive then there exists $x \in X$ such that $xPx$.

\item
\label{lem:transitivity-02}
\emph{(SCT criterion)} If $(S,P)(S,P)=(S,P)$ then there exists $x \in X$ such that $xPx$.
\end{enumerate}
\end{lemma}

\begin{proof}
\begin{enumerate}
\item
By the hypothesis that $\Graph$ satisfies PDC, there exist $x,y \in \dom{S \cup P}$ such that $xPy$ and with more reason $x(S \cup P)y$. On the other hand, since $y \in \dom{S \cup P}$, there is $z \in X$ such that $y(S \cup P)z$. From this, by transitivity, it follows that
$x(S \cup P)z$, hence $x = y$ since $(S,P)$ is injective. From $x=y$ and $xPy$ we conclude
that $xPx$.
\item
$(S,P)(S,P)=(S,P)$ implies that $(S,P)$ is transitive. Now we apply the previous point.
\end{enumerate}
\end{proof}

Using Theorem 4 of \cite{Lee1} from PDC we may already deduce the SCT criterion,
then conclude SCT. We prefer to include a 
more direct proof, 
using the infinite Ramsey's Theorem. We preliminarily introduce some terminology for it. 
$[N]^2 = \Set{\Set{x,y} \subseteq N \mid x \neq y}$ is the set of edges of the complete undirected graph whose vertices are in $N$. 

A {\em finite coloring} of $[N]^2$ is a map  
$c:[N]^2 \to \underline{k}$ for some $k > 0$, where $\underline{k} = \Set{1, \ldots , k}$, the set of ``colors'', is the set of the first $k$ positive integers.
A subset $[A]^2 \subseteq [N]^2$ is {\em monochromatic} w.r.t. the coloring $c$ if there is $i \in \underline{k}$ such that  $c((a,b)) = i$ for all
$(a,b) \in [A]^2$.

\begin{theorem}[Ramsey's Theorem]\label{thr:Ramsey}
For all finite coloring $c$ of $[\Nat]^2$ there exists $A \subseteq \Nat$, $A$ infinite, such that $[A]^2$ is monocromatic with respect to $c$.
\end{theorem}

\begin{proof} See e.g. Theorem 5 in Chapter 1 of \cite{Graham1980}.
\end{proof}

By abuse of language, we say that $c:E \to \underline{k}$ is a finite coloring for any set $E$ of edges.

\begin{corollary}\label{cor:Ramsey}
Let $G = (V, E)$ be a complete undirected graph with infinite denumerable $V$, and let $c: E \to \underline{k}$ be a finite coloring of its edges,
then there exists an infinite $U \subseteq V$ that is a complete subgraph of $G$ and such that for all distinct $u, v \in H$ we have $c(\Set{u,v}) = i$ for some $i \in \underline{k}$.
\end{corollary}

\begin{proof}
As $V$ is denumerable, the graphs $[\Nat]^2$ and $[V]^2$ are isomorphic.
Hence,  up to isomorphism, the corollary is a particular case of Theorem \ref{thr:Ramsey}.
\end{proof}

\begin{theorem}[Main Theorem]\label{thr:PDC+injImpliesGTC}
Let $\Graph = (V, \biRel, E)$ be a transition graph with injective $\biRel$; if $\Graph$ satisfies PDC, then it satisfies SCT.
\end{theorem}

\begin{proof}
Let $\ppath : v_0 \stackrel{(S_0,P_0)}{\longrightarrow} v_1 \stackrel{(S_1,P_1)}{\longrightarrow} v_2 \stackrel{(S_2,P_2)}{\longrightarrow}\cdots$
be any infinite path of $\Graph^* = (V, \biRel^*, E^*)$, where $v_i = \source{\ppath(i)}$ for $i < \omega$.
For all $0 \leq i < j < \omega$ there is an edge $\edgestar{(S_{i,j}, P_{i,j})}{v_i}{v_j}$, 
defined by
$(S_{i,j}, P_{i,j}) = (S_{i}, P_{i}) \comp \cdots \comp (S_{j-1}, P_{j-1})$.

Let $G = (\mathcal{V}, \mathcal{E})$ be the undirected graph formed by the sets
$\mathcal{V} = \Set{(i, v_i) \mid i < \omega}$, which is denumerable, and $\mathcal{E} = \Set{\Set{(i, v_i), (j, v_j)} \mid i < j}$; hence $G$ is a complete graph.

The edges in $\mathcal{E}$ correspond
to the edges $\edgestar{(S_{i,j}, P_{i,j})}{v_i}{v_j}$ of $\Graph^*$. 
Since $(S_{i,j}, P_{i,j}) \in \biRel^*$, which is a finite set, the mapping
$c\Set{(i, v_i), (j, v_j)} = (S_{i,j}, P_{i,j})$ is a finite coloring of $\mathcal{E}$.

By Corollary \ref{cor:Ramsey}, there is an infinite $\mathcal{H} \subseteq \mathcal{V}$ such that for some $(S, P) \in \biRel^*$ and for all $i, j \in \mathcal{H}$, $i \not  = j$ we have
$c\Set{(i, v_i), (j, v_j)} = (S, P)$. 

As $V$ is finite, there exists $u \in V$ such that for infinitely many $i < \omega$
we have $(i, u) \in \mathcal{H}$.
This determines a cofinal subpath $\ppath_{\mathcal{H}} : u \stackrel{(S,P)}{\longrightarrow} u \stackrel{(S,P)}{\longrightarrow} u \cdots$ of $\ppath$ where $(S,P)$ is self-composing and, by the above, the \quotationMarks{colors} $(S,P)\comp (S,P)$ and $ (S,P)$ are the same.
By the hypotheses that $\Graph$ is
injective and satisfies PDC, Lemma \ref{lem:transitivity}.\ref{lem:transitivity-02} 
implies that there exists $x \in X$ such that $x P x$.  
It remains to show that there is a trace $\tau$ such that $\tau \models \ppath$ and that, whenever $v_i = u$, $\tau(i) = x$.
Let $i_0 < i_1$ be the first two indexes such that $v_{i_0} = u = v_{i_1}$ and $(S,P) = (S_{i_0,i_1},P_{i_0,i_1})$; then in $\ppath$ we have the subpath:
\[u = v_{i_0} \stackrel{(S_{i_0},P_{i_0})}{\longrightarrow} \cdots \stackrel{(S_{i_1},P_{i_1})}{\longrightarrow} v_{i_1} = u \]
where we assume that the $v_j \neq u$ for all $i_0 < j < i_1$. 

Since $x P x$ and clearly $P \subseteq S \cup P = (S_{i_0} \cup P_{i_0}) \cdots (S_{i_1-1} \cup P_{i_1-1})$, there is a sequence $x = x_{i_0}, \ldots , x_{i_1} =x$ such that for all $i_{0} \leq i < i_1$, $x_i(S_{i} \cup P_{i}) x_{i+1}$ 
and $x_i P_{i} x_{i+1}$ for some $i_{0} \leq i < i_1$.
Assume that $i_0 < i_1 < i_2 < \ldots$ are the indexes such that 
$v_{i_0} = v_{i_1} = v_{i_2} = \ldots = u$.
Now, by repeating the same construction for all consecutive indexes $i_k, i_{k+1}$ in the
sub-sequence, 
 we obtain a trace $\tau$ such that $\tau \models \ppath$, $\tau(i) = x$ for infinitely many $i$, 
 including infinitely many progress points w.r.t. $P$. 
 Hence $\Graph$ satisfies SCT by the arbitrary choice of $\ppath$.
\end{proof}

\section{Related Work and Conclusions}
\label{sec:Related}

The sequent calculus with inductive definitions in Section \ref{sec:cyclic} is a variant of the systems in \cite{Brotherston1,Brotherston2} and \cite{BrotherstonSimpson}, to which it can be shown to be equivalent in the case the system includes Peano Arithmetic.
We omit such a proof in this paper for space reasons. 
In the literature, the global trace condition is defined in terms of infinitely progressing traces or as infinite descending sequences. They are both instances of Fermat's method of infinite descent, which is used to argue by contradiction that a cyclic pre-proof satisfying the GTC is semantically valid. It is known that valid cyclic proofs correspond to \emph{finite} proofs with induction if the system includes arithmetic, while in some other cases they prove more theorems \cite{Berardi}.

The transition graphs are a variant of size-change graphs in \cite{Lee1,Lee2,Lee3}; in particular, injective transition graphs correspond to fan-in free size-change graphs
as defined in \cite[\S 2.4]{Lee3}. Size-change graphs are a technique for verifying program termination by building suitable abstractions of functional programs so that the SCT condition implies termination. 

In Ikebuchi's \cite{Mirai}, it is shown that, up to an appropriate association of a set of size-change graphs such that the composition determines the paths in the proof, the GTC and SCT conditions are equivalent, see there, Proposition 2. This result is rephrased for Transition Graphs 
in our Proposition \ref{prop:GTC-SCT}. 

To the best of our knowledge, the Progress Domain Condition, presented here in Section 
\ref{sec:progressDomainCond}, is a novel contribution.
It is decidable by the finiteness of the bi-partite relations involved in a transition graph, and it is general enough, as we have shown in Theorem \ref{thr:injective-pre-proofs},
in the case that the transition graphs associated with 
$\CLJIDomega$ and $\CLKIDomega$ are injective.

The proof of Theorem \ref{thr:PDC+injImpliesGTC} is non-constructive since it is based on infinite Ramsey's Theorem, which is a purely classical result. However, we know
of a constructive proof of this theorem \cite{BERARDI20151382}, which we have omitted for the sake of brevity. We understand Theorem \ref{thr:PDC+injImpliesGTC} and its constructive proof as a step toward efficient translations of cyclic proofs into ordinary inductive proofs, with an intuitionistic
translation in the case of intuitionistic cyclic proofs. These translations, in turn, are motivated
by cross-checking formal cyclic proofs using proof assistants that do not support cyclic proofs.
For the moment, such translations exist only in theory and are built over the
results of \cite{Berardi} and \cite{Mirai}.

In \cite{Lee1}, it is proved that the SCT problem in the general case is PSPACE-complete. 
From  \cite{Lee3}, restricting to fan-in free size-change graphs, a lower bound of SCT
for known algorithm is exponential in $n$, where $n$ essentially depends on the size of the set of formal parameters of the analyzed program (more precisely on the size of expressions defining ranking functions), which corresponds to the set of the occurrences of inductive predicates in a cyclic pre-proof. A brute force algorithm deciding PDC is also exponential, but the relation between the complexity of PDC and the lower bound of SCT based on the size of ranking functions 
deserves further study.

\newpage

\bibliography{references}

\end{document}